\RequirePackage{lineno}
\documentclass[aps,prl,onecolumn,amsmath,amsart,amssymb,11pt,superscriptaddress]{revtex4}
\usepackage{graphicx}
\usepackage[T1]{fontenc}
\usepackage{amsmath}
\usepackage{color}
\usepackage{times}

\usepackage{ulem}

\begin{document}
%\setpagewiselinenumbers
%\modulolinenumbers[1]
%\linenumbers

\title{Spinon Fermi surface spin liquid in a triangular lattice antiferromagnet NaYbSe$_2$}

\author{Peng-Ling Dai$^\#$}
\affiliation{Center for Advanced Quantum Studies and Department of Physics, Beijing Normal University, Beijing 100875, China}

\author{Gaoning Zhang$^\#$}
\affiliation{School of Physical Science and Technology, ShanghaiTech University, Shanghai 201210, China}

\author{Yaofeng Xie}
\author{Chunruo Duan}
\affiliation{Department of Physics and Astronomy, Rice University, Houston, TX 77005, USA}

\author{Yonghao Gao}
\affiliation{State Key Laboratory of Surface Physics, Department of Physics,
Fudan University, Shanghai 200433, China}

\author{Zihao Zhu}
\affiliation{State Key Laboratory of Surface Physics, Department of Physics,
Fudan University, Shanghai 200433, China}

\author{Erxi Feng}
\affiliation{Neutron Scattering Division, Oak Ridge National Laboratory, Oak Ridge, TN 37831}

\author{Zhen Tao}
\affiliation{Center for Advanced Quantum Studies and Department of Physics, Beijing Normal University, Beijing 100875, China}

\author{Chien-Lung Huang}
\affiliation{Department of Physics and Astronomy, Rice University, Houston, TX 77005, USA}

\author{Huibo Cao}
\author{Andrey Podlesnyak}
\author{Garrett E. Granroth}
\author{Michelle S. Everett}
\author{Joerg C. Neuefeind}
\affiliation{Neutron Scattering Division, Oak Ridge National Laboratory, Oak Ridge, TN 37831}

\author{David Voneshen}
\affiliation{ISIS Facility, Rutherford Appleton Laboratory, Chilton, Didcot, Oxfordshire OX11 0QX, UK}
\affiliation{Department of Physics, Royal Holloway University of London, Egham, TW20 0EX, UK}

\author{Shun Wang}
\affiliation{School of Physics, Huazhong University of Science and Technology, Wuhan 430074, China}

\author{Guotai Tan}
\affiliation{Center for Advanced Quantum Studies and Department of Physics, Beijing Normal University, Beijing 100875, China}

\author{Emilia Morosan}
\affiliation{Department of Physics and Astronomy, Rice University, Houston, TX 77005, USA}

\author{Xia Wang}
\affiliation{School of Physical Science and Technology, ShanghaiTech University, Shanghai 201210, China}

\author{Hai-Qing Lin}
\affiliation{Beijing Computational Science Research Center, Beijing 100094, China}

\author{Lei Shu}
\affiliation{State Key Laboratory of Surface Physics, Department of Physics,
Fudan University, Shanghai 200433, China}

\author{Gang Chen}
\email{gangchen.physics@gmail.com}
\affiliation{Department of Physics and HKU-UCAS Joint Institute 
for Theoretical and Computational Physics at Hong Kong, The University of Hong Kong, Hong Kong, China}
\affiliation{State Key Laboratory of Surface Physics, Department of Physics,
Fudan University, Shanghai 200433, China}

\author{Yanfeng Guo}
\email{guoyf@shanghaitech.edu.cn}
\affiliation{School of Physical Science and Technology, ShanghaiTech University, Shanghai 201210, China}

\author{Xingye Lu}
\email{luxy@bnu.edu.cn}
\affiliation{Center for Advanced Quantum Studies and Department of Physics, Beijing Normal University, Beijing 100875, China}

\author{Pengcheng Dai}
\email{pdai@rice.edu}
\affiliation{Department of Physics and Astronomy, Rice University, Houston, TX 77005, USA}

\date{\today}

\begin{abstract}
Triangular lattice of rare-earth ions with interacting effective spin-$1/2$ 
local moments is an ideal platform to explore the physics of quantum spin 
liquids (QSLs) in the presence of strong spin-orbit coupling, crystal electric fields,  
and geometrical frustration. The Yb delafossites, NaYbCh$_2$ (Ch=O, S, Se) with 
Yb ions forming a perfect triangular lattice, have been suggested to be candidates 
for QSLs. Previous thermodynamics, nuclear magnetic resonance, and 
powder-sample neutron scattering
measurements on NaYbCh$_2$ have supported the suggestion of the QSL ground states. 
The key signature of a QSL, the spin excitation continuum, arising from 
the spin quantum number fractionalization, has not been observed. Here 
we perform both elastic and inelastic neutron scattering measurements 
as well as detailed thermodynamic measurements on high-quality single 
crystalline NaYbSe$_2$ samples to confirm the absence of long-range magnetic order
down to 40 mK, and further reveal a clear signature of magnetic excitation continuum
extending from 0.1 to 2.5 meV. The comparison between the structure of the magnetic excitation 
spectra and the theoretical expectation from the spinon continuum suggests that the ground state of NaYbSe$_2$ is a QSL with a spinon Fermi surface.
\end{abstract}

\maketitle

${\bf Introduction.}$ 
The quantum spin liquid (QSL) is a correlated quantum state 
in a solid where the spins of the unpaired electrons are 
highly entangled over long distances, yet they do not 
exhibit any long-range magnetic order in the zero temperature limit.  
Originally proposed by Anderson as
the ground state for a system of $S=1/2$ spins on 
a two-dimensional (2D) triangular lattice that interact 
antiferromagnetically with their nearest neighbors \cite{anderson1973}, a
QSL is a novel quantum state of matter beyond the traditional Landau's symmetry
breaking paradigm \cite{balents2010,zhou2017,balents2017,broholm2020}, 
and might be relevant for our understanding of high-temperature superconductivity \cite{anderson1987,rmp2006,Lee2007} and quantum computation in certain cases \cite{kitaev2003,kitaev2006}.  
Beyond the simple characterization of absence of a magnetic order, 
one key signature of the excitations in a QSL is the presence of deconfined spinons 
that are fractionalized quasiparticles carrying spin-1/2,  
observed by inelastic neutron scattering as a spin excitation
continuum fundamentally different from the integer  
spin-wave excitations 
in an ordered magnet \cite{Han2012,jun2016,mourigal2017,ysli17, ysli19, Balz2016,Gao2019,Gaudet2019}.

Although spin excitation continuum has been observed in the geometrically frustrated  
$S=1/2$ single crystal systems with 2D Kagom\'{e} \cite{Han2012},
2D triangular \cite{jun2016,mourigal2017}, three-dimensional (3D) 
distorted Kagom\'{e} bilayers \cite{Balz2016}, and 3D pyrochlore~\cite{Gao2019,Gaudet2019} 
lattices, there is no consensus on the microscopic origin
of the observed spin-excitation continuum.  In the 2D $S=1/2$ Kagom\'{e} 
lattice ZnCu$_3$(OD)$_6$Cl$_2$ \cite{Han2012} and an effective $S=1/2$ 
triangular-lattice magnet YbMgGaO$_4$ \cite{ysli2015_1, ysli2015_2, jun2016,mourigal2017, ysli17, ysli19}, different 
interpretation of the observed spin-excitation continuum includes 
a spin glass state of magnetic \cite{Freedman2010} and 
nonmagnetic Mg-Ga site disorder due to intrinsic sample issues~\cite{Ma2018,Zhu2017,kimchi2018}, 
respectively, rather than the fractionalized quasiparticles of a QSL~\cite{broholm2020}. 
To conclusively identify the presence of deconfined spinon excitations in a QSL, 
one needs to search for the expected spin-excitation continuum among 
candidate QSL materials with high quality single crystals and establish their 
physical properties with clear experimental signatures and structures.

Recently, geometrically frustrated 2D triangular-lattice 
rare-earth-based materials with effective ${S=1/2}$
local moments have attracted considerable attentions \cite{Rau2018,Maksimov2019}. 
Compared with YbMgGaO$_4$~\cite{ysli2015_1},
the family of Yb dichalcogenide delafossites NaYbCh$_2$ (Ch=O, S, Se) 
does not have the issue of Mg-Ga site disorders in the non-magnetic 
layers and thus provides a genuine example for an 
interacting spin-1/2 triangular-lattice antiferromagnet~\cite{qmzhang2018,baenitz2018,ranjith_o}. Moreover, NaYbCh$_2$ exhibit larger magnetic anisotropy than YbMgGaO$_4$ \cite{Hozoi2019, ESR_1, ESR_2, qmzhang2020, jun2016}, suggesting that the in-plane magnetic interactions play the dominant role. 
The combination of the strong spin-orbit coupling (SOC) and the crystal electric field (CEF)
leads to a Kramers doublet ground state for the Yb$^{3+}$ ion in NaYbCh$_2$ that gives rise to
 the effective $S=1/2$ local moment at each ion site. 
Since the energy gaps between the ground and first excited Kramers doublets 
CEF levels for NaYbSe$_2$ [Fig.~1(b)] \cite{qmzhang2020}, NaYbS$_2$ \cite{baenitz2018}, 
and NaYbO$_2$ \cite{ranjith_o} are well above $\sim$12 meV, the magnetic properties
below 100 K can be safely interpreted from the interaction between the effective $S=1/2$ 
local moments.
Although previous experiments on powder samples of NaYbO$_2$ provided 
some positive evidence for QSL ground states \cite{wilson2019,ding2019, wilson2020},
there are no detailed neutron scattering experiments on single crystalline 
samples to establish the presence of the magnetic excitation continuum and 
further reveal its wave vector, energy, temperature dependence, which are essential for identifying the nature of the possible QSL state.  Here we report magnetic susceptibility, 
heat capacity, and neutron scattering results on single crystals of NaYbSe$_2$.
In addition to confirming the absence of magnetic order
down to 40 mK and spin freezing down to 90 mK, we show the presence of a 
spin excitation continuum extending from 0.1 to 2.5 meV. 
Since our careful structure refinement and pair-distribution function (PDF) analysis reveal 
only $\sim5\%$ of Yb on Na site and no evidence for a 
spin glass state at 40 mK, we conclude that 
the ground state of NaYbSe$_2$ has signatures of a QSL,
consistent with the expectation of a spinon 
Fermi surface QSL state \cite{SI, LiY2017}.

${\bf Results.}$ High-quality single crystals of NaYbSe$_2$ were grown by using flux method with Te as the flux (see Methods for further synthesis and experimental details). Figure 1(a) displays schematics of crystal structure and reciprocal space of NaYbSe$_2$, where Yb ions form a perfect triangular lattice layer. Inelastic neutron scattering spectra of CEF excitations obtained by subtracting the scattering of a non-magnetic reference NaYSe$_2$ from NaYbSe$_2$ are shown in Fig.~1(b) \cite{SI}. Consistent with previous work \cite{qmzhang2020}, the CEF levels of Yb$^{3+}$ have a Kramers doublet ground state and three excited Kramers doublets at $E=15.7$, 24.5, and 30.2 meV at $T=7$ K, thus ensuring that all
measurements below $\sim100$ K can be safely considered as an effective $S=1/2$ ground state \cite{qmzhang2020}. Note that the CEF levels with $E=24.5$ and 30.2 meV
are instrumental energy resolution limited \cite{SI} while the lowest CEF excitation is broader than the energy resolution indicative of the internal structure of this mode. The broadening could be attributed to the Yb$^{3+}$-Yb$^{3+}$ exchange effects on CEF excitations that can split the lowest CEF excitation into two levels \cite{wilson2020, SI}. 

To characterize the behavior of the local moments of Yb and their exchange interactions, we measured the magnetic susceptibility of single-crystalline NaYbSe$_2$. The temperature dependence of magnetization and the in-plane magnetic susceptibility $\chi_{\perp}(T)$ is depicted in Fig.~1(c), and a simple fit to the 
Curie-Weiss law yields $\Theta_{\rm CW,\perp}\simeq -13$ K in the low-temperature region ($<20$ K), whose absolute value is larger than $|\Theta_{\rm CW,\perp}|\simeq 7$ K when the Van Vleck contribution is subtracted \cite{ranjith_se}, indicating the predominantly antiferromagnetic spin interactions in NaYbSe$_2$. Heat capacity measurements were also performed to characterize the thermodynamics of NaYbSe$_2$, and the magnetic contribution $C_{\rm mag}(T)$ to the specific heat of NaYbSe$_2$ and its dependence on applied magnetic fields from $0$ T to $8$ T are presented in Fig.~1(d). The data shows a broad peak that shifts upward in temperature as a function of increasing magnetic field for H $\parallel c$, no sharp anomaly indicative of the onset of long-range order, consistent with the susceptibility result and earlier work \cite{ranjith_se}. Figure 1(e) also shows the estimated temperature dependence of $C_{\rm mag}(T)/T$ (left axis) and the corresponding magnetic entropy $S_{\rm mag}$ (right axis). It is noted that $C_{\rm mag}(T)/T$ in the low-temperature regime ($<0.5$ K) is almost a constant, well compatible with the fact that the spinon Fermi surface alone has a constant density of states and would give a heat capacity depending linearly on temperature. Moreover, the temperature dependence of the magnetic entropy saturates to a value close to $S_{\rm mag}\approx R\ln2$ (where $R$ is the ideal gas constant) around 15 K, consistent with an effective spin-$1/2$ description of the Yb$^{3+}$ local moment~\cite{ranjith_se}. 

Although stoichiometric NaYbSe$_2$ has no intrinsic structural disorder in the Na${^+}$ intercalating layer  \cite{qmzhang2018,baenitz2018,ranjith_o}, real crystal could still have structural defects in Na$^+$ and Se$^{2-}$ sites, and these vacant sites could be replaced by Yb$^{3+}$ and Te$^{2-}$, respectively (see Methods). 
To accurately determine the stoichiometry of our NaYbSe$_2$, we carried out single crystal 
X-ray structural refinement by recording 1334 Bragg reflections, corresponding to 238 non-equivalent reflections. The Rietveld refinement results of the single-crystal X-ray diffraction data collected at $T=250$ K are shown in Fig.~1(f) and the fitting outcome reveals full occupancy of the Yb$^{3+}$ ($3a$) and Se$^{2-}$ ($6c$) sites in the YbSe$_2$ layers and $\sim4.8\%\pm1\%$ of the Na ($3b$) sites occupied by the Yb ions. A small amount of Yb occupying the Na site is not surprising because Yb and Na have the same chemical environment for bonding, where both cations have six Se coordinates and are located at the center of NaSe$_6$/YbSe$_6$ octahedra \cite{Hozoi2019}. To further characterize the structural character of the sample, we have also performed PDF analysis on neutron diffraction data measured on 2.7 grams of NaYbSe$_2$ powder ground from large amount of single crystals obtained from the same batches as the spin-excitation measurements. As shown in Fig. 1(g), the local PDF peaks are well reproduced by fitting with the refined average structure using the X-ray diffraction data, indicating the absence of substantial local distortions. The average structure includes a Yb substitution at the Na site and possible excess Te at the Se site. 
The PDF analysis suggests an upper limit of 10\% of Yb at the Na site and 0\% Te at the Se site. While this value is larger than that obtained by single crystal X-ray refinements, single crystal refinement results are more accurate as more Bragg peaks are measured in the X-ray refinements. These results are consistent with the inductively-coupled plasma measurements of chemical composition of the sample (see Methods for details). Although Yb ions in the Na layers may be magnetic, our  frequency-dependent ac susceptibility measurements down to 90 mK can be well described with a Curie-Weiss fit and 
 show no evidence of spin freezing [Fig. 1(h)].

In the previous inelastic neutron scattering measurements on single crystals of 
CsYbSe$_2$ ($\Theta_{\rm CW}\simeq -13$ K), spin excitations were found to be centered 
around the $K$ point in reciprocal space [Fig.~1(a)], with no intensity modulation along
the $c$-axis, and extending up to 1 meV \cite{xing2019}. To determine what happens in
NaYbSe$_2$, we must first determine if the system has long/short-range magnetic order.
For this purpose, we align the crystals in the $[H,H,0]\times [0,0,L]$ and 
$[H,0,0]\times [0,K,0]$ zones [Fig. 1(a)]. 
Figures 2(a) and 2(b) display maps of elastic scattering in the 
$[H,H,L]$ and $[H,K,0]$ planes, respectively, at $T=40$ mK (top panels) 
and $40$ mK$-10$ K (bottom panels).  In both cases, no evidence of long/short range 
magnetic order was observed at 40 mK, consistent with previous 
magnetic susceptibility, heat capacity, and nuclear magnetic
resonance measurements \cite{ranjith_se}. The wave vector dependence of 
the spin excitations of $E=0.3\pm 0.1$ meV
in the $[H,H,L]$ zone at 40 mK (left panel) and 10 K (right panel) 
is presented in Fig.~2(c). At 40 mK, one can see a featureless rod 
of scattering along the $[1/3,1/3,L]$ direction, 
indicating that spin excitations in NaYbSe$_2$ are 2D in nature and have no $c$-axis modulations.  The scattering becomes much weaker at 10 K, thus confirming the magnetic nature of the scattering at 40 mK. Moreover, Fig.~2(d) shows the temperature dependence of the $E=0.3\pm 0.1$ meV spin excitations in the $[H,K,0]$ zone. The magnetic scattering is centered around the $K$ point, consistent with the previous work \cite{xing2019}, and decreases significantly with increasing temperature.  

To further reveal the intrinsic quantum dynamics of the local moments of the Yb ions, 
we perform the inelastic neutron scattering measurements to study the spin excitations 
in single crystals of NaYbSe$_2$ at both 40 mK and 10 K. Constant-energy images of spin excitations 
with a variety of energies in the in-plane 2D Brillouin zones at 40 mK and 10 K are summarized 
in Figs.~3(a-d) and 3(e-h), respectively. At $E=0.15\pm 0.05$ meV and 40 mK, the magnetic scattering 
spectral weights spread broadly 
in the Brillouin zone but with higher intensity at the $K$ point and no scattering near the zone center 
(the $\Gamma$ point) [Fig. 3(a)]. This is clearly different from the wave vector dependence of the low-energy magnetic scattering for YbMgGaO$_4$, in which the spectral weight is enhanced around the $M$ point \cite{jun2016}. 
The high intensity at the $K$ point in NaYbSe$_2$ might arise from the strong $XY$-type exchange interaction, 
since the strong SOC in this material indeed brings certain anisotropic interactions \cite{gang2017}. 
With increasing energies to $E=0.6\pm 0.1$ [Fig. 3(b)], $1.1\pm 1$ [Fig. 3(c)],
and $E=2.1\pm 0.1$ meV [Fig. 3(d)], the magnetic 
scattering spectral weights become more evenly distributed in the Brillouin zone
and gradually decrease with increasing energy. While the spin-excitation continuum at $E=0.15\pm 0.05$ meV nearly vanishes on 
warming from 40 mK to 10 K [Fig. 3(e)], 
the spectral weights at other energies become weaker but are still located around 
the Brillouin zone boundaries, especially the scattering at the $K$ points [Figs. 3(f-g)].

Figures 4(c) and 4(d) display the wave vector-energy dependence 
of the spin excitation spectral intensity (in log scale) 
along the magenta color arrow direction
in Fig. 4(a) at 40 mK and 10 K, respectively. The excitation continuum here is analogous to that calculated from the free spinon theory. The excitation bandwidth ($\sim 2$ meV), together with the Curie-Weiss temperatures, characterize the scale of the magnetic interactions \cite{jun2016, LiY2017}. At both 40 mK and 10 K, the spectral intensity is broadly distributed in the energy-momentum plane, and the excitation intensity gradually decreases with increasing energy and finally vanishes above  $\sim$2.2 meV. The broad neutron-scattering spectral intensity at 40 mK persists to the lowest energy that we measured implying a high density of spinon scattering states at low energies. Moreover, the spectral weight around $\Gamma$ point is suppressed to form a V-shaped upper bound \cite{SI}. Combining these two facts, it strongly suggests a spinon Fermi surface QSL since this scenario not only provides a high density of spinon states near the Fermi surface, but also well explains the V-shaped upper bound on the excitation energy near the $\Gamma$ point~\cite{LiY2017}. 

The V-shaped structure is one of the key properties for the magnetic excitation in the spinon Fermi surface QSL. It arises from the large density of states and the linear $E-k$ spinon dispersion near the Fermi surface. Due to the spin quantum number fractionalization, the neutron scattering creates the spinon particle-hole pairs across the spinon Fermi surface. To excite the pair with an energy $E$, a minimal momentum transfer $E/v_F$ is needed where $v_F$ is the Fermi velocity. The slope of the V-shape is expected to be the Fermi velocity. It is also noted that the low-energy spin excitations clearly peak around the $K$ point at 40 mK [Fig. 4(c)], and they decrease dramatically on warming but still keep the V-shaped upper bound around $\Gamma$ point at 10 K [Fig. 4(d)]. In addition,  Figs.~4(e) and 4(f) present the wave vector-energy dependence of the spin-excitation spectral intensity along the magenta color arrow directions in Fig. 4(b) at 40 mK and 10 K, respectively. The main results are similar to that in Figs. 4(c) and 4(d), and also support a spinon Fermi surface QSL.

The data points in Figs. 5(a) and 5(b) show energy dependence of spin excitations 
at the $K_1$ and $M_2$ points, 
respectively, under a variety of temperatures $T=40$ mK, 2 K, and 10 K. 
The solid lines in 
the figures display similar data at the $\Gamma_1$ point. 
Consistent with Fig.~4, magnetic scattering clearly 
decreases with increasing temperature at the 
$K_1$and $M_2$ points, and essentially vanishes at the $\Gamma_1$ point. 
The temperature differences (40 mK$-$10 K) of the imaginary part
of the dynamic susceptibility, $\chi^{\prime\prime}(E)$, at the 
$K_1$ and $M_2$ points peak around 0.15 and 0.3 meV, respectively, as shown in the 
inset in Fig.~5(b). Besides, Fig.~5(c) compares energy dependence of the magnetic
scattering at the $M_1$, $M_2$, and $K_1$ with the background at the $\Gamma_2$ point. 
To show the wave vector dependence of 
spin excitations, Figs.~5(d-g) plot the spectral intensity along the $[H,H,0]$ direction 
for various energies of $E=0.25\pm 0.1$, $0.5\pm 0.1$, $1.3\pm 0.1$, 
and $2.3\pm 0.1$ meV, respectively, at $T=40$ mK, 2 K, and 10 K. Similarly, 
Figs.~5(h) and 5(i) also plot constant-energy cuts along the $[0.5-K,0.5+K,0]$ 
direction for energies of $E=0.3\pm 0.1$, $0.9\pm 0.1$, $1.5\pm 0.1$, $2.3\pm 0.1$ meV at 40 mK and 10 K, respectively. All the results are compatible with Figs. 4(c-f). In Figs. 5(a)-5(b), the spin excitations can only be resolved above $E\sim 0.15$ meV because of the instrumental energy resolution. To further check whether the excitations are gapless, we show in Fig. 6 spin-excitation energy spectra at $K_1$ point measured with improved instrumental energy resolution ($E_i=1.55$ meV). The energy dependent spin excitations for $T=40$ mK and $10$ K reveals the persistence of the spin excitations down to $E\sim0.06$  meV, indicative of the gapless nature for the excitations \cite{SI}.

{\bf Discussion and Conclusion} 
Overall, the magnetic and heat capacity measurements, combined with the neutron scattering results on single crystals of NaYbSe$_2$ demonstrate the absence of long/short-range magnetic order even down to 40 mK, implying a quantum disordered QSL state. In particular, besides the naive disorder and the simple spectral continuum of spin excitation, the almost linear temperature dependence of magnetic heat capacity $C_{\rm mag}(T)$ at the low temperature regime, the enormous low energy gapless excitations and the V-shaped upper bound around the $\Gamma$ point in inelastic neutron scattering spectrum all strongly indicate the existence of a spinon Fermi surface. Theoretically, although the pure compact $U(1)$ gauge theory in two spatial dimensions is always confined due to the non-perturbative instanton events~\cite{Polyakov1977}, it has been shown and understood that in the presence of spinon Fermi surface and gapless excitations, the QSL phase could be stable against gauge fluctuations, and a noncompact $U(1)$ gauge theory remains to be a good low energy description~\cite{SSLee2008,Lee2007}. Therefore, our experimental results and conclusion about spinon Fermi surface QSL can be compatible with theory.
%In the neutron scattering study of YbMgGaO$_4$, Li {\it et al.} interpreted the high-energy excitations ($\gtrapprox0.5$meV) as the breaking of the nearest valence bonds \cite{ysli17}, and the low-energy excitations as the propagation of unpaired spins via the rearrangement of valence bonds. As this interpretation does not refer to specific type of QSL, it could also apply here and is not against with the spinon Fermi surface picture. 
The scenario of spinon Fermi surface QSL could further be verified by low-temperature thermal-transport measurement, which has an advantage to unveil the nature of low-energy itinerant excitations.      

Very recently, the pressure-induced insulator to metal transition followed by an emergence of superconductivity in NaYbSe$_2$ was observed in experiments~\cite{Jia2020, Zhang2021}. This is quite remarkable since the QSL has long been thought to be a parent state of the high-temperature superconductivity \cite{anderson1987,rmp2006,Lee2007}. It was suggested that doping a QSL could naturally result in superconductivity~\cite{anderson1987,rmp2006,Lee2007} due to the intimate relationship between high-temperature superconductivity and QSL, but the definitive experimental evidence showing that doping QSLs give rise to superconductivity is still lacking. Instead of doping, Ref.~\cite{Jia2020, Zhang2021} obtained the superconductivity by pressure, which opens up a promising way to study the superconductivity in QSL candidates and sheds light on the mechanism of high temperature superconductivity.  

{\bf Acknowledgments}

We thank M. Stone for suggestions of appropriate neutron scattering instrumentation, Feng Ye (ORNL) for the assistance with the single-crystal x-ray diffraction measurements. We also thank Ling Wang and Yuesheng Li for helpful discussions.
The research at Beijing Normal University is supported by the National Natural Science Foundation of China (Grant No. 11734002 and 11922402). The work at ShanghaiTech university is supported by the National Natural Science Foundation of China (No. 11874264, Y.G.). Y.G. and X.W. thank the support from Analytical Instrumentation Center (\# SPST-AIC10112914), SPST, ShanghaiTech University. 
The neutron scattering work at Rice is supported by US DOE BES DE-SC0012311 (P.D.).
This work is further supported by funds from the Ministry of Science and Technology of China ( 
grant No.2016YFA0301001, No.2018YFGH000095, No.2016YFA0300500 for G.C., and No.2016YFA0300501, No.2016YFA0300503 for L.S. and G.C.)
and from the Research Grants Council of Hong Kong with General 
Research Fund Grant No.17303819 (G.C.). 
E.F. and H.C. acknowledges support of US DOE BES Early Career Award KC0402010 under Contract DE-AC05-00OR22725.
E.M. and C.-L.H. acknowledge support from US DOE BES DE-SC0019503.
This research used resources at Spallation Neutron Source, a DOE Office of Science User Facility operated by ORNL. We gratefully acknowledge the Science and Technology Facilities Council (STFC) for access to neutron beamtime at ISIS.

\begin{figure*}[tbh]
\includegraphics[width=15cm]{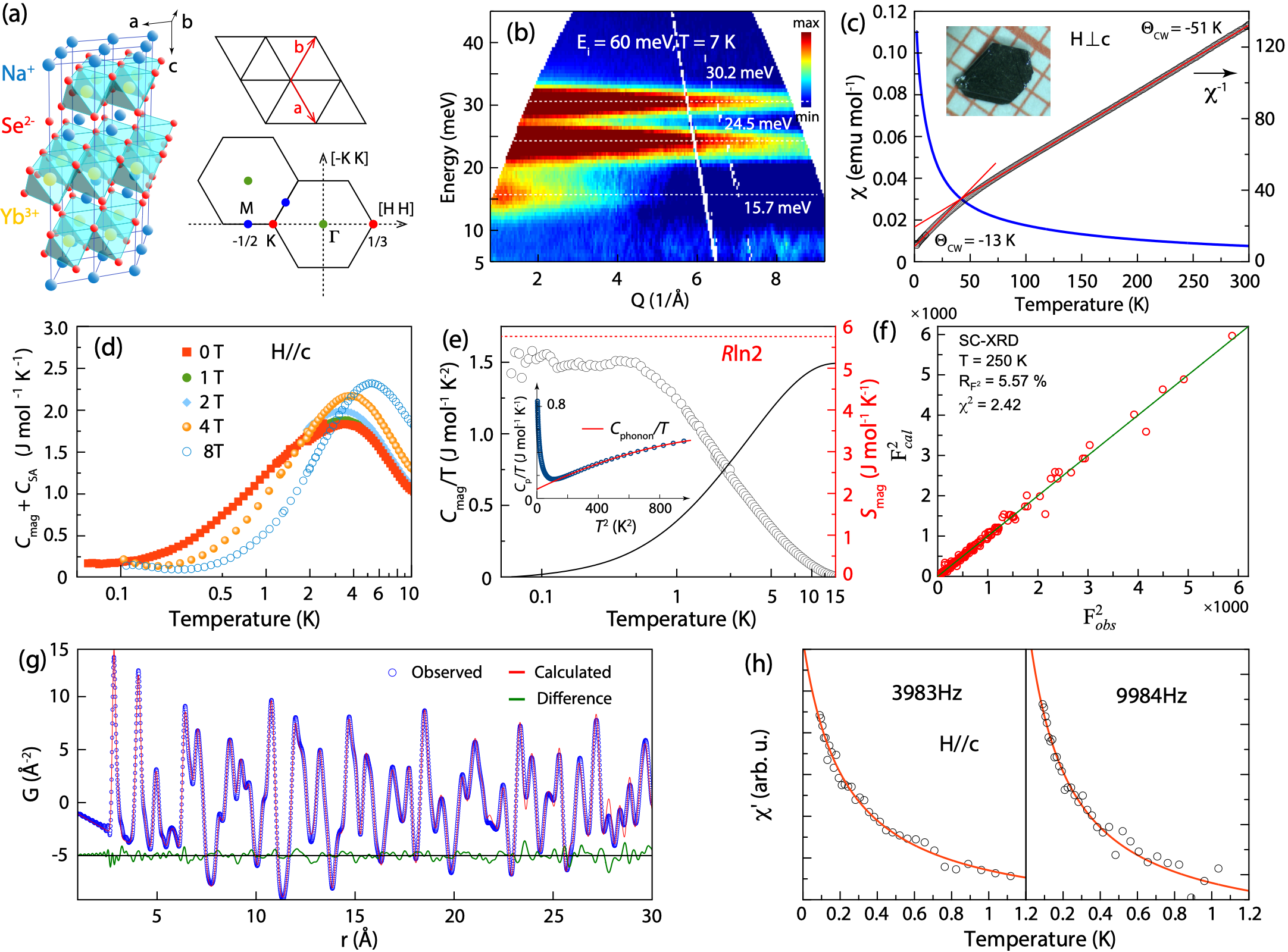}
\caption {{\bf Crystal structure and reciprocal space, CEF levels, heat capacity and stoichiometry of
NaYbSe$_2$}.
(a) The structure of NaYbSe$_2$ and corresponding reciprocal space. The lattice parameters are $a=b\approx4.07$ {\AA}, $c\approx20.77$ {\AA}. (b) Inelastic neutron scattering spectra of CEF excitations obtained by subtracting the
scattering of a non-magnetic reference NaYSe$_2$ from the intensity of NaYbSe$_2$ (see Methods), in which the energy axis of NaYSe$_2$ was shifted by -1.7 meV for calibrating the difference in phonon energies \cite{SI}. Three CEF energy levels are marked by white dashed lines. (c) Temperature-dependent magnetic susceptibility along $H\perp c$ direction. The fitting $\frac{1}{\chi-\chi_0}=\big(\frac{C}{T-\Theta_{\rm CW,\perp}}\big)^{-1}$ for high-temperature range ($\sim160-300$K) results in a Curie-Weiss temperature $\Theta_{\rm CW,\perp}\approx-51$K, and the low temperature range ($<20$K) generates a $\Theta_{\rm CW,\perp}\approx-13$ K, in which $C$ is Curie constant, and $\chi_0\sim2\times10^{-4}$ emu/mol is a temperature-independent background term. The inset shows the crystal for the magnetization measurements. (d) Temperature dependent magnetic contribution ($C_{\rm mag}$) to the specific heat (with minor contribution from nuclear Schottky anomaly $C_{\rm SA}$) of NaYbSe$_2$ and its dependence on applied magnetic fields $H\parallel c$ \cite{SI}. Phonon contribution has been subtracted. (e) Temperature dependent $C_{\rm mag}/T$ (black circle) with $C_{\rm SA}/T$ subtracted \cite{SI} and the magnetic entropy (black curve). The red dashed line marks the value of $R\ln 2$. The inset shows $C_p/T$ as a function of $T^2$. The red solid curve is a fitting of the phonon contribution $C_{\rm phonon}$. (f) The Rietveld refinement results of the single-crystal X-ray diffraction data at 250 K yield Na$_{0.952(10)}$Yb$_{0.048(10)}$YbSe$_2$.  F$^2 _{cal}$ and F$^2_{obs}$ are the calculated and observed structure factors, respectively. (g) The PDF analysis of neutron data on NaYbSe$_2$ up to 30 \AA. The weighted residual value is 9.56\%. (h) AC susceptibility of NaYbSe$_2$ single crystal measured with frequencies of 3983 Hz and 9984 Hz. The red solid curves are Curie-Weiss fits for the data.} 
\label{fig1}
\end{figure*}

\begin{figure*}[tbh]
\includegraphics[width=10cm]{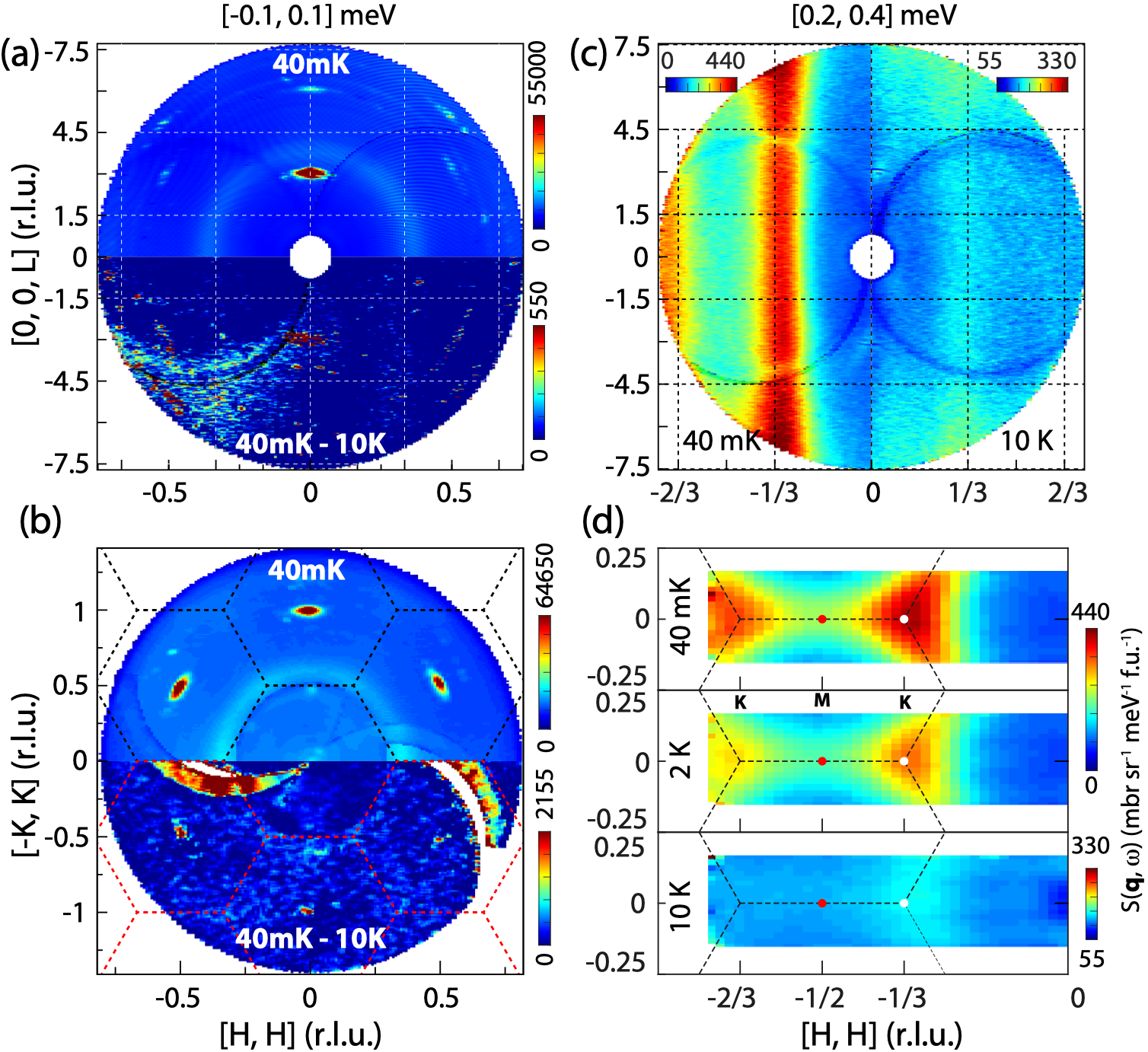}
\caption {{\bf Neutron  scattering results in $[H,H,L]$ and $[H,K,0]$ zones}. Elastic neutron scattering results ($E=0\pm0.1$ meV) in (a) the $[H,H,L]$ plane and (b) $[H,K,0]$ plane measured with $E_i=3.32$ meV and $3.70$ meV, respectively. Scattering along the vertical direction ($[-K,K,0]$ for (a) and $[0, 0, L]$ for (b)) is integrated. The upper half panels of (a) and (b) are data at $T=40$ mK, and the lower are the differences between $T = 40$ mK and 10 K. (c) $L$-dependence of the spin excitations along the $[H, H]$direction at $T=40$ mK (left half panel) and $T=10$ K (right half panel), with $K=[-0.05, 0.05]$ and $E=0.3\pm0.1$ meV. (d) Spin excitations with $E=0.3\pm0.1$ in the $[H, K]$ plane measured at $T=40$ mK, $2$, and $10$ K. Scattering along the $[0, 0, L]$ direction is  integrated. The black dashed lines mark the Brillouin zones of NaYbSe$_2$.
}
\label{fig2}
\end{figure*}

\begin{figure*}[tbh]
\includegraphics[width=16cm]{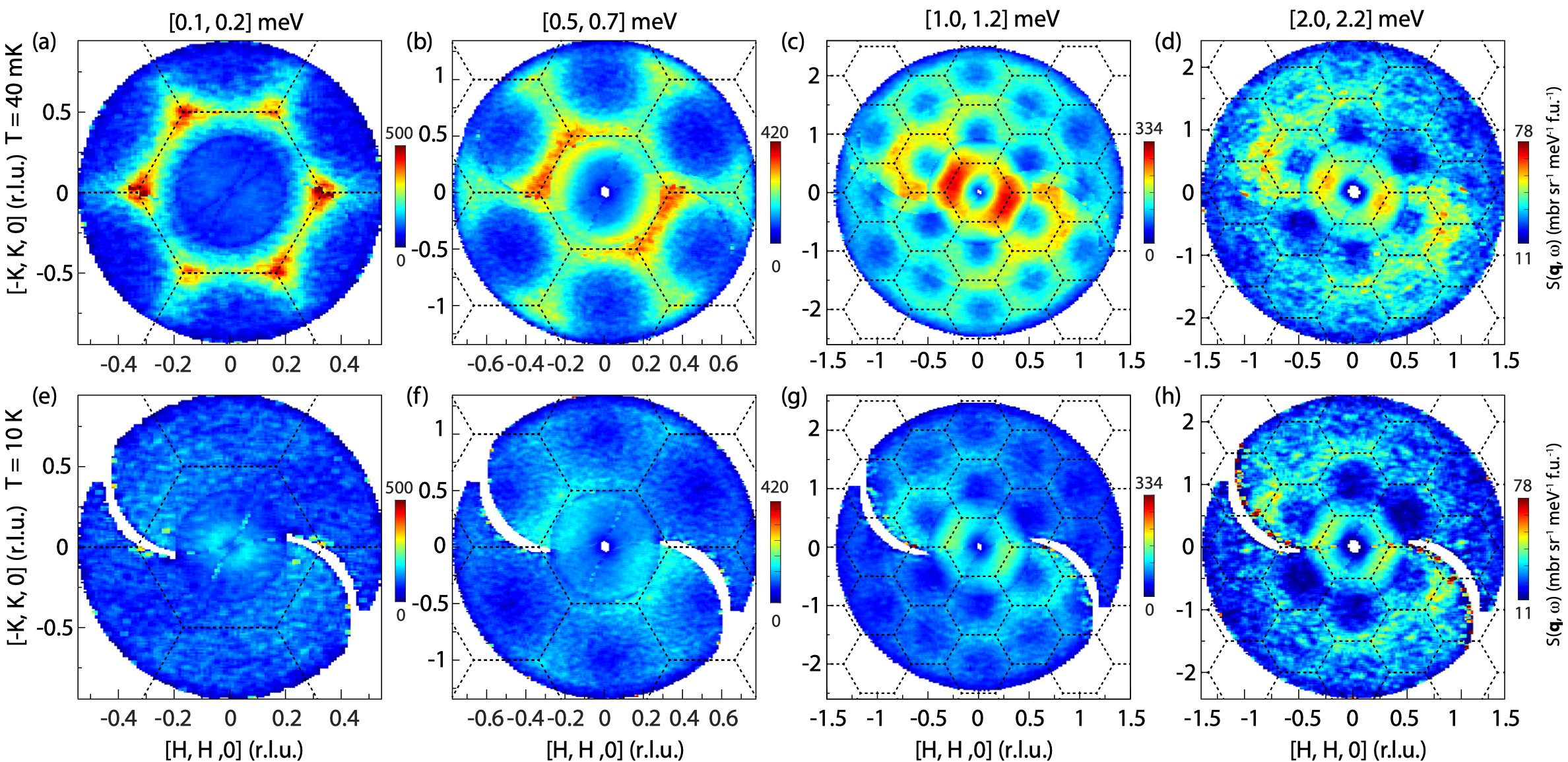}
\caption {
{\bf Constant-energy images of spin excitations in the $[H, K, 0]$ plane}.
(a-d) Images at $T=40$ mK and (e-h) $10$ K. The intensity along the vertical $[0, 0, L]$direction is integrated. Spin excitations for (a,e) $E=0.15\pm0.05$, (b,f) $0.6\pm0.1$, (c,g) $1.1\pm0.1$, and (d,h) $2.1\pm0.1$ meV are measured with $E_i=1.77$, $3.70$, $12.14$ and $12.14$ meV, respectively. The black dashed lines mark the Brillouin zones in the reciprocal space. The data are collected in $180^{\circ}$ range of sample rotation around the $c$-axis. The $360^{\circ}$ circular coverage are generated by averaging the raw data and its mirror in the $[H, K, 0]$ plane. The C$_2$-like anisotropy has been attributed to a trivial effect caused by sample-volume change in beam during sample rotation for neutron scattering measurements in [H, K, 0] plane \cite{SI}.
}
\label{fig3}
\end{figure*}

\begin{figure*}[tbh]
\includegraphics[width=16cm]{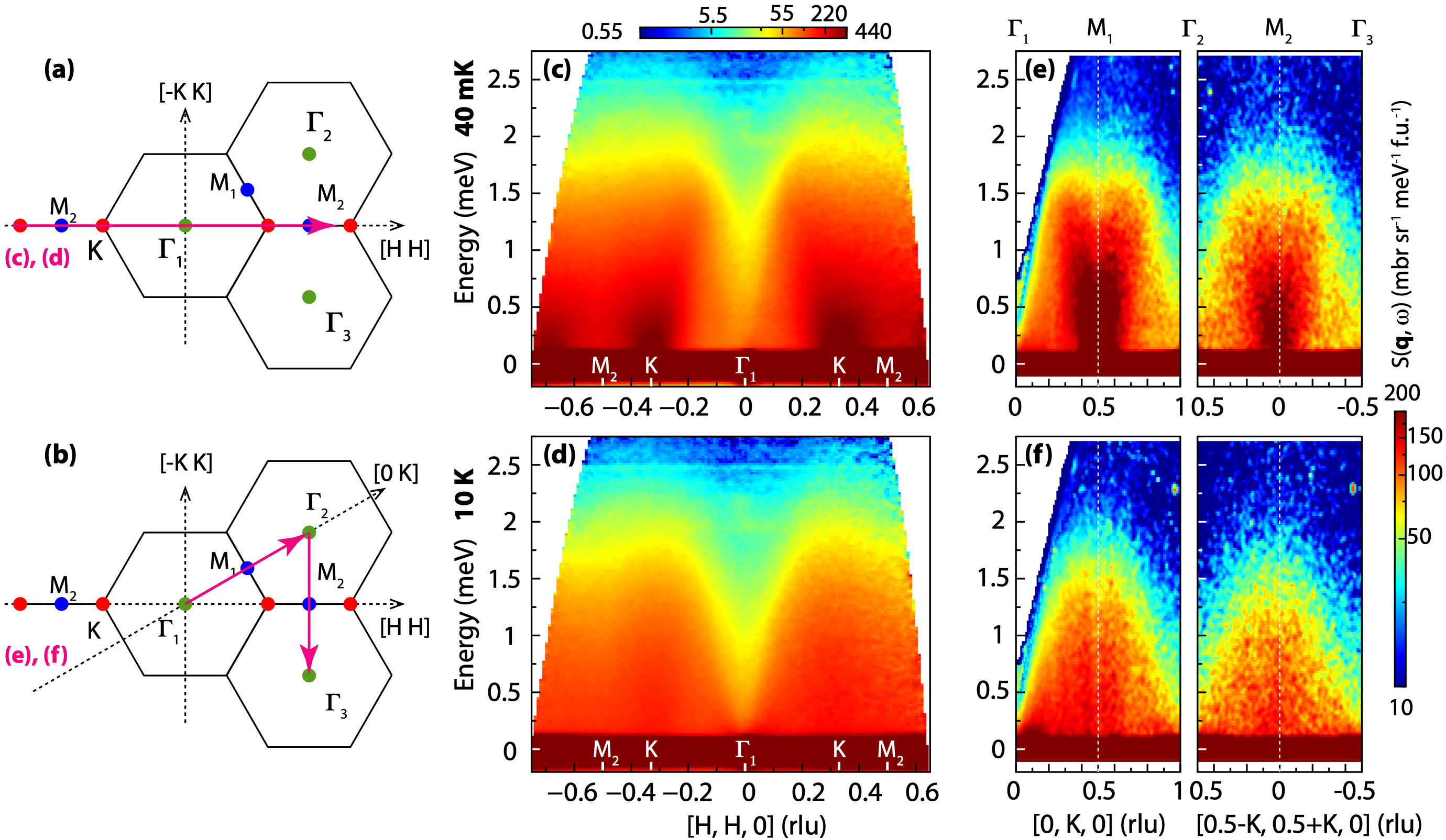}
\caption {{\bf Spin excitation spectra along high symmetry momentum directions.} 
(a,b) Schematics of the Brillouin zones with high symmetry points $\Gamma$, $K$,
 and $M$ denoted by green, red, and blue dots, and the high symmetry directions for the images in (c-f) marked by pink lines with arrow heads. Spin excitation spectra collected at (c) $T=40$ mK and (d) 10 K along the $M_2$-$K$-$\Gamma$-$K$-$M_2$ with $E_i=3.32$ meV. (e,f) Intensity color maps along the $\Gamma_1$-$M_1$-$\Gamma_2$ and $\Gamma_2$-$M_2$-$\Gamma_3$ directions measured with $E_i=3.7$ meV. 
}
\label{fig4}
\end{figure*}

\begin{figure*}[tbh]
\includegraphics[width=16cm]{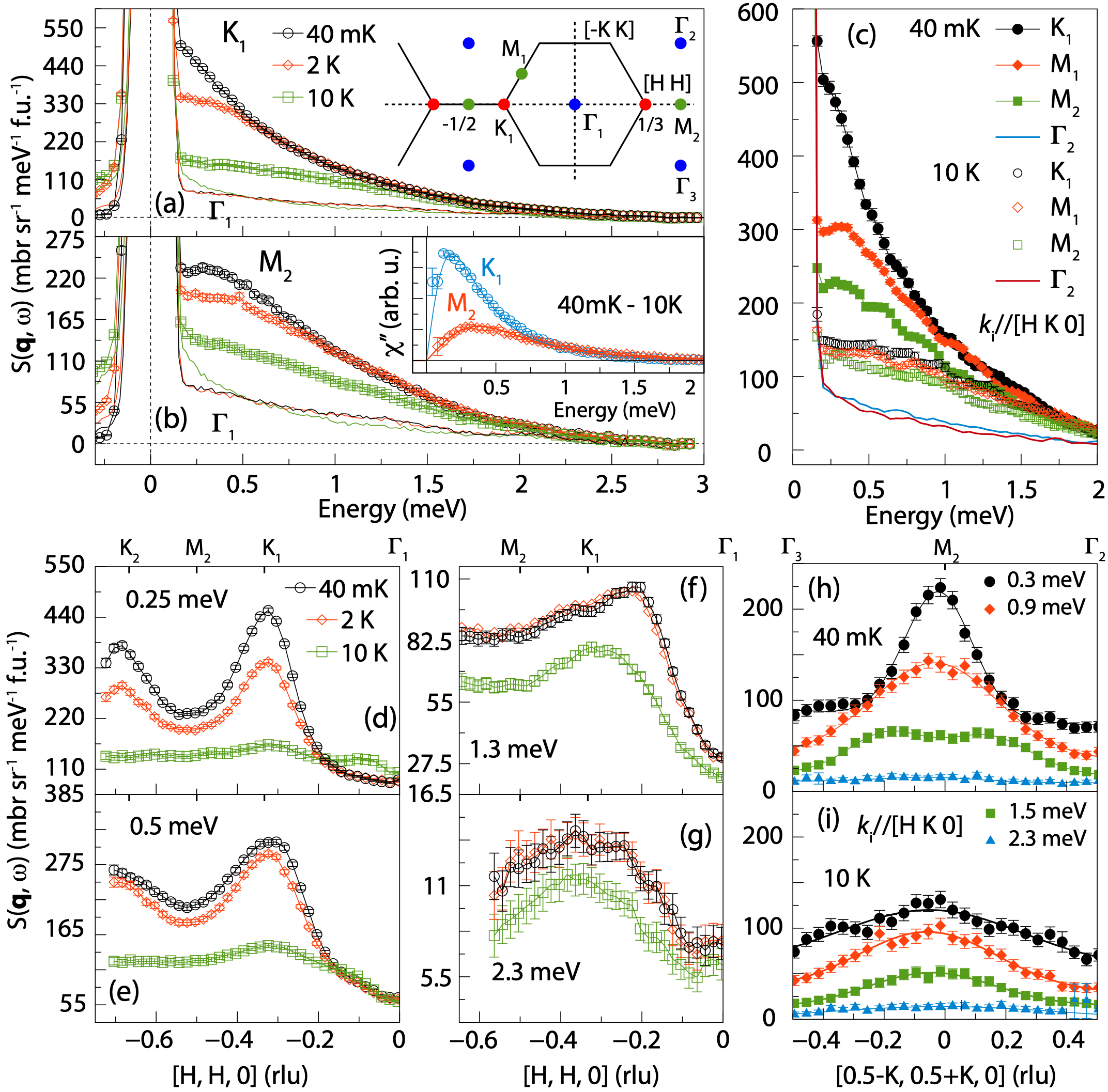}
\caption {{\bf Wave vector dependence of spin excitations along high symmetry directions}.
The wave vector cuts in (a,b,d-g) were measured in the $[H,H,L]$ zone with $E_i=3.32$ meV, while those in 
(c,h,i) were measured in the $[H,K,0]$ plane with $E_i=3.70$ meV. (a) and (b) show the energy dependent scattering at ${K_1}$ and ${M_1}$ points measured at $T = 40$ mK (black circle), 2 K (red diamond) and 10 K (green square). The inset in (a) is a schematic of the reciprocal space with the ${\Gamma}$, ${K}$ and ${M}$ points denoted by green, red and blue dots. The black, red, and green curves are energy cuts at $\Gamma_1$. The inset of (b) shows the difference of $\chi''$ between the spectra for $T=40$ mK and $10$ K at the ${K}$ and ${M}$ points. The light blue and red curves are fittings of the $\chi^{\prime\prime}$ with a damped harmonic oscillator model. (c) shows the energy cuts at the $K_1$, $M_1$, $M_2$ and $\Gamma_2$. Solid symbols represent the data collected at $T=40$ mK and the open symbols collected at 10 K. The black and blue curves are energy cuts at the $\Gamma_2$ point  measured at $T=40$ mK and 10 K. (d-g) Constant energy cuts along the $M_2$-$K_1$-$\Gamma_1$ for $T=40$ mK, 2 K, and 10 K, with corresponding energy transfers marked in the panels. 
Constant energy cuts along the $\Gamma_3$-$M_2$-$\Gamma_2$ measured at (h) $T=40$ mK and (i) 10K. The solid curves are guides to the eyes and the error bars represent one standard deviation.
}
\label{fig5}
\end{figure*}

\begin{figure*}[tbh]
\includegraphics[width=9cm]{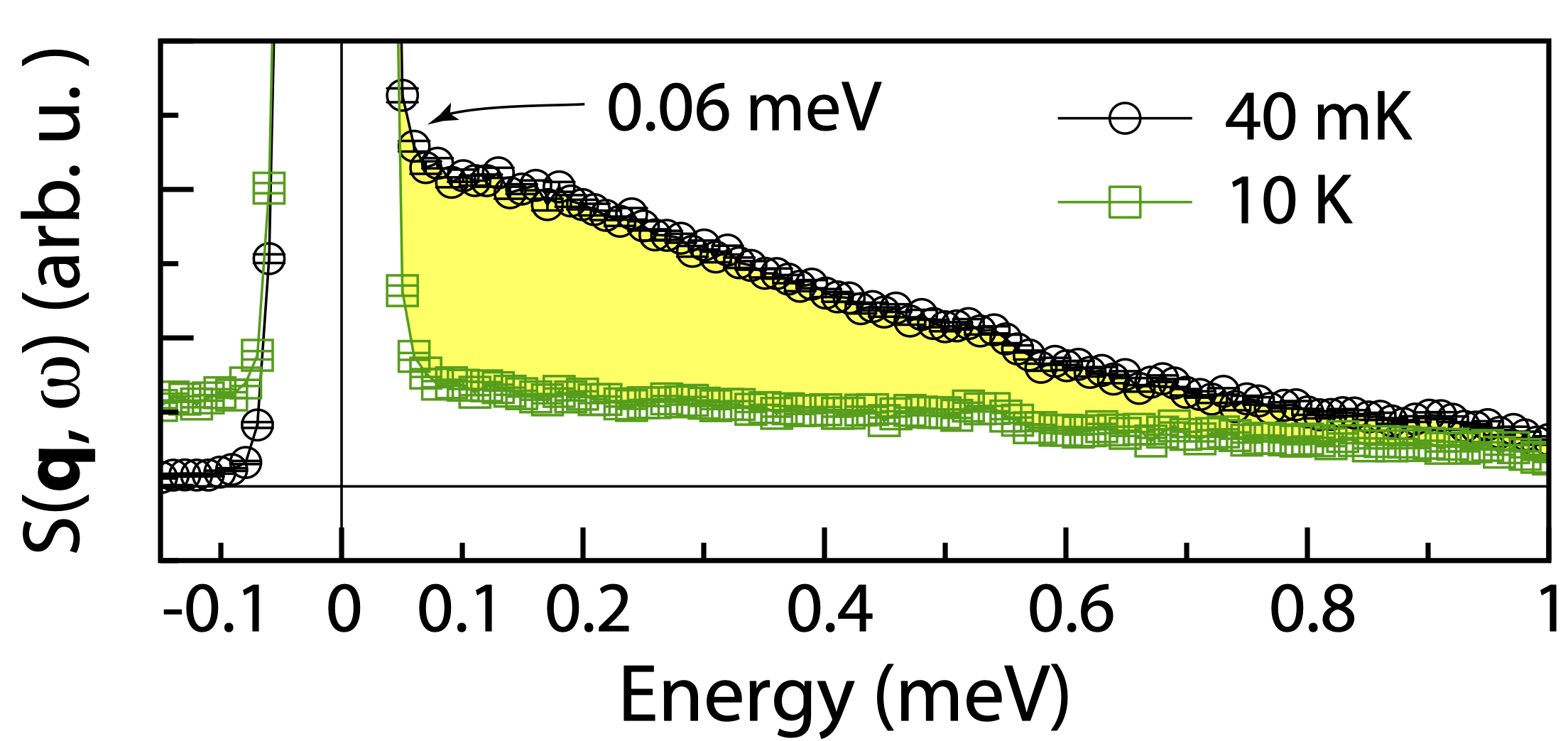}
\caption {Spin excitation energy spectra at $K_1$ position measured with $E_i=1.55$ meV at $T = 40$ mK (red circles) and 10 K (green squares). The yellow shaded area marks the difference between the spectra for $T = 40$ mK and $10$ K. The black arrow marks the lowest energy (0.06 meV) magnetic excitations.}
\label{fig6}
\end{figure*}

\section{Methods}

{\bf Crystal Growth}
All the NaYbSe$_2$ single crystals used in this study were grown by using Te as the flux. The starting materials are in a molar ration of Na : Yb : Se : Te = 1 : 1 : 2 : 20. To avoid the violent reaction between Na and Se, the Na (99.7\%) blocks and Te (99.999\%) granules were mixed and slowly heated up to 200$^\circ$C within 20 hours and pre-reacted at the temperature for 10 hours. The precursor was then thoroughly mixed with Yb (99.9\%) blocks and Se (99.999\%) granules in the molar ratio and placed into an alumina crucible. The crucible was sealed into a quartz tube under the vacuum of $10^{-4}$ Pa and then slowly heated up to 950$^\circ$C within 15 hours. After the reaction at this temperature for 20 hours, the assembly was slowly cooled down to 800$^\circ$C at a temperature decreasing rate of 1$^\circ$C/h. At 800$^\circ$C, the quartz tube was immediately taken out of the furnace and placed into a high-speed centrifuge to separate the excess Te flux. To show a comparison, NaYbSe$_2$ crystals were also grown by using NaCl as the flux in the similar procedure as mentioned above (not used for this study) \cite{SI}. The crystallographic phase and quality of the grown crystals were examined on a Bruker D8 VENTURE single crystal X-ray diffractometer using Mo K$_{\alpha1}$ radiation ($\lambda$ = 0.71073{\AA}) at room temperature. The crystals grown by using different flux have the same high quality \cite{SI}. Growth of the polycrystalline NaYbSe$_2$ and NaYSe$_2$ samples has been described elsewhere \cite{ranjith_se}.

{\bf Stoichiometric Analysis}
The single crystal X-ray diffraction of NaYbSe$_2$ were performed at 250 K on Rigaku XtaLAB PRO diffractometer at Spallation Neutron Source, ORNL. Structure refinement based on the X-ray diffraction data were carried out with FullProf suite \cite{fullprof}, generating (Na$_{0.952(10)}$Yb$_{0.048(10)}$)YbSe$_2$ without Te occupying Se sites. Elemental analysis of a group of NaYbSe$_2$ single crystals grown with Te flux with a total mass of 35mg were performed by inductively-coupled plasma (ICP) method on Thermo Fisher ICP 7400 system. The result---Na$_{0.965}$Yb$_{1.03}$Se$_{1.98}$Te$_{0.025}$---can be interpreted as $\sim 3\%$ of Na$^+$ sites being occupied by Yb ions and agrees well with the structure refinement results of single-crystal x-ray diffraction, especially considering that Te could exist as flux in the sample.

{\bf Heat Capacity} The specific heat capacity of NaYbSe$_2$ was measured down to 50 mK using a thermal-relaxation method in DynaCool-PPMS (Physical Property Measurement System, Quantum Design) with the magnetic field applied along the $c$-axis at Fudan University and Rice University. The total specific heat is described as a sum of magnetic and lattice contributions: $C_p = C_{\rm mag} + C_{\rm phonon}$. We fit the phonon contribution with $C_{\rm phonon}=\beta T^3 + \alpha T^5$.

{\bf Magnetic Susceptibility} The magnetic susceptibility of a rare-earth magnetic system with strong spin-orbit coupling can be determined by CEF excitations, particularly the first CEF excitation level. In this case, the Curie-Weiss analysis is applicable only in the limited temperature range much smaller than the first CEF excitation level. In NaYbSe$_2$, spin-orbit coupling is quite a large energy scale and generates the local moment $J=7/2$ for Yb$^{3+}$ ion. The crystal field further splits the eight $J=7/2$ states into 4 Kramers doublets, and the ground state doublet contributes to the effective spin-1/2 description that is responsible for the low temperature magnetism. Since the lowest CEF excitation is $\sim15.7$meV ($\sim180$ K), the Curie-Weiss fitting of the $T$ < 20 K range is not affected by spin-orbit coupling and CEF levels.

{\bf Neutron Scattering}
The neutron scattering measurements of the magnetic excitations in [H, H, L] scattering plane, and the CEF excitations were performed on the Cold-Neutron-Chopper-Sepctrometer (CNCS) \cite{cncs} and ARCS \cite{arcs} at the Spallation Neutron Source (SNS), Oak-Ridge National Laboratory (ORNL), respectively. The measurements in [H, K, 0] scattering plane were carried out on the LET cold neutron chopper spectrometer \cite{let, let_data}, ISIS spallation neutron source, Rutherford Appleton Laboratory (RAL), UK. We co-aligned $\sim 3.7$ grams of NaYbSe$_2$ single crystals for the measurements of magnetic excitations and prepared $\sim 10$ grams NaYbSe$_2$ and NaYSe$_2$ polycrystalline samples for the CEF excitation measurements. The powder neutron diffraction experiment for pair-distribution function analysis were performed at NOMAD, ORNL at 100 K, with $2.7$ grams of NaYbSe$_2$ polycrystalline sample ground from $\sim 100$ pieces of single crystals obtained from the same batches as the 3.7 gram sample set for our elastic/inelastic neutron scattering experiments at CNCS and LET. The neutron scattering data was reduced with Mantid \cite{mantid} and analyzed with MantidPlot, Horace \cite{horace}, and MSlice.

\end{document}